\documentclass[twocolumn,noshowpacs,preprintnumbers,amsmath,amssymb]{revtex4}

\usepackage{verbatim}

\usepackage{graphicx}% Include figure files
\usepackage{dcolumn}% Align table columns on decimal point
\usepackage{bm}% bold math

\begin{document}

\title{Field-effect transistors assembled from functionalized carbon nanotubes}
\author{Christian Klinke}
\email{klinke@chemie.uni-hamburg.de}
\altaffiliation{Present address: Institute of Physical Chemistry, Universtity of Hamburg, 20146 Hamburg, Germany.}
\affiliation{IBM T. J. Watson Research Center, 1101 Kitchawan Road, Yorktown Heights, NY 10598, USA}

\author{James B. Hannon}
\affiliation{IBM T. J. Watson Research Center, 1101 Kitchawan Road, Yorktown Heights, NY 10598, USA}

\author{Ali Afzali}
\affiliation{IBM T. J. Watson Research Center, 1101 Kitchawan Road, Yorktown Heights, NY 10598, USA}

\author{Phaedon Avouris}
\affiliation{IBM T. J. Watson Research Center, 1101 Kitchawan Road, Yorktown Heights, NY 10598, USA}

\begin{abstract} % abstract goes here

We have fabricated field effect transistors from carbon nanotubes using a novel selective placement scheme. We use carbon nanotubes that are covalently bound to molecules containing hydroxamic acid functionality. The functionalized nanotubes bind strongly to basic metal oxide surfaces, but not to silicon dioxide. Upon annealing, the functionalization is removed, restoring the electronic properties of the nanotubes. The devices we have fabricated show excellent electrical characteristics.

\end{abstract}

\maketitle

One-dimensional nanostructures, such as nanowires~\cite{C01} and carbon nanotubes~\cite{C02,C03} (CNTs) have excellent electrical properties making them attractive for applications in nanotechnology~\cite{C04,C05,C06,C07}. Semiconducting nanotubes, in particular, are receiving considerable attention due to their very promising performance as channels of field effect transistors (FETs)~\cite{C08,C09,C10,C11,C12}. However, in order for the large-scale integration of CNTs in electronics to take place, many challenges must be overcome. One particularly important problem is selectively positioning CNTs on a substrate. Although considerable progress has been made for nanoclusters~\cite{C13,C14,C15} and nanowires~\cite{C16,C17,C18} it remains a difficult task for CNTs. Several approaches have been demonstrated based on the chemical patterning of the substrate (typically SiO2)~\cite{C19,C20,C21}. In these approaches surfaces are patterned with molecules that either promote or prevent nanotube adhesion. For example, amine-terminated compounds have been shown to enhance nanotube binding while hydrophobic compounds prevent it~\cite{C22}. Nevertheless, these methods remain unsatisfactory because of the adverse influence of the functionalization on the electrical performance of the devices. 

\begin{figure}[!h]
\begin{center}
\includegraphics[width=0.45\textwidth]{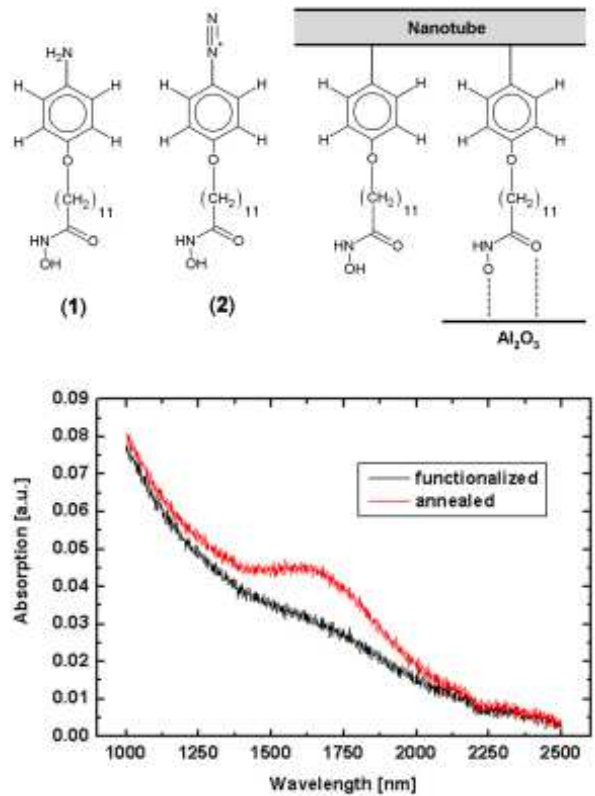}
\caption{\it Top: Schematic of the nanotube functionalization procedure and the adsorption on Al2O3. Bottom: Optical absorption spectra of nanotubes covalently functionalized with AMUHA. Black: functionalized, red: annealed at 600$^{\circ}$C for 120~s in N2. The annealed nanotubes show the characteristic E11 absorption peak.}
\label{F01}
\end{center}
\end{figure}

Here we propose a new approach that allows to selectively position CNTs based on their functionalization rather than the modification of the substrate.  The complete fabrication sequence of chemical functionalization, selective placement, fabrication of multiple devices and defunctionalization is demostrated. High-performance devices can thus be directly assembled. Specifically, we covalently bound molecules to the CNTs that contain the hydroxamic acid functionality (Fig.~\ref{F01}). The functionalized tubes bind strongly to Al2O3 (and other basic metal oxides), but not to SiO2. Upon heating to 600$^{\circ}$C, the functionalization is removed, and the electrical properties of the tube are restored. We exploited the selective bonding to position nanotubes at precise locations on a SiO2 substrate patterned with Al2O3. Using this approach we fabricated FETs with high yield and excellent electrical properties. 

%\begin{wrapfigure}{r}{0.4\textwidth}
%\end{wrapfigure}

Our functionalization strategy is based on the reaction of CNTs with aryldiazonium salts, first demonstrated by Bahr et al.~\cite{C23} This reaction which occurs at room temperature has been used extensively to attach various organic compounds to CNTs~\cite{C24,C25,C26,C27}. For the selective placement of nanotubes on metal oxides we combine it with the hydroxamic acid functionality. The acid groups react with hydroxyl groups on the surfaces of the metal oxides; alkylhydroxamic acids are known to form strong bonds with native oxides of metals~\cite{C28}. To achieve these two requirements, we designed and synthesized 11-(4-aminophenoxy)-1-undecylhydroxamic acid (AMUHA) \textbf{(1)}. Reaction of AMUHA with nitrosonium tetrafluoroborate in acetonitrile generated the corresponding diazonium salt \textbf{(2)} which without isolation was added to a dispersion of laser-ablation CNTs in aqueous sodium dodecyl sulphate (SDS). This functionalized the CNTs covalently by forming a bond between the nanotube and the molecule without destroying the lattice structure of the nanotubes. The CNTs were isolated by addition of large amounts of acetone and then centrifugation. The tubes were then dispersed in either methanol or ethanol. The functionalized nanotubes remain in solution for several weeks.

We characterized the functionalized tubes using optical spectroscopy, electron microscopy, and electrical transport measurements. The one-dimensional character of CNTs leads to the formation of strongly bound exciton states. The lowest allowed exciton occurs in the infra-red, and is easy to detect using adsorption spectroscopy. The spectrum recorded from a dense sample of covalently AMUHA functionalized tubes (dried on quartz) is featureless, indicating that the band structure of the functionalized tube is very different from that of a clean nanotube (Fig.~\ref{F01}). After annealing at 600$^{\circ}$C for 120~s in N2 (ramp 5$^{\circ}$/s) a broad adsorption band around 1650~nm appears, associated with the first dipole active exciton (E11)~\cite{C29,C30,C31}. The broad shape of the absorption peak is due to the presence of a distribution of nanotubes in the sample, with diameters ranging from 0.8 to 1.2~nm~\cite{C32}. The appearance of this band upon annealing indicates that the one-dimensional band structure of the nanotube is restored, most likely because the covalent bonds between the nanotubes and the AMUHA are broken~\cite{C33}. The spectra obtained after annealing are similar to those obtained for pristine nanotube samples.

%\begin{wrapfigure}{r}{0.4\textwidth}
\begin{figure}[!h]
\begin{center}
\includegraphics[width=0.45\textwidth]{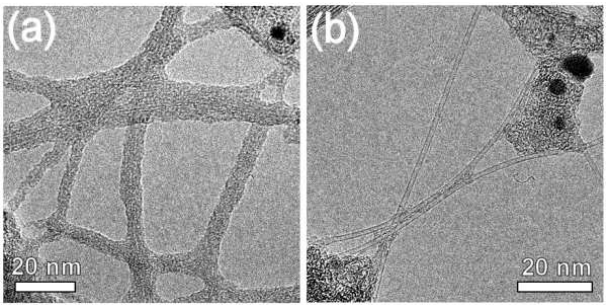}
\caption{\it TEM micrographs of (a) functionalized and (b) annealed nanotubes. The annealed nanotubes lost the functionalization and exhibit remarkably clean surfaces.}
\label{F02}
\end{center}
\end{figure}
%\end{wrapfigure}

Transmission electron microscopy (TEM) and atomic force microscopy (AFM) measurements show that we have produced nanotubes that are uniformly functionalized. TEM measurements were made by depositing a film of functionalized tubes dispersed in methanol on ``holey'' SiO2 coated TEM grids. Before annealing, the TEM micrographs show a dense, amorphous layer covering the tubes (Fig.~\ref{F02}a). However, after annealing to 600$^{\circ}$C for 120~s in N2, the amorphous layer disappears (Fig.~\ref{F02}b). No defects are seen in TEM and the tubes appear remarkably clean (e.g. free of amorphous carbon). The structural integrity of the nanotubes is maintained during the annealing process.

AFM measurements show that annealing functionalized nanotubes dispersed on SiO2 substrates also leads to a complete removal of the functionalization. Before anneal, individual tubes (determined from the apparent height in AFM) have an average diameter of about 1.8~nm. Measurements along the length of the tubes confirm that the tubes are uniformly functionalized. Following annealing to 600$^{\circ}$C for 120~s in N2 the average diameter is reduced to about 1.0~nm. No residual contamination of the nanotube surface is observed. 

%\begin{wrapfigure}{r}{0.4\textwidth}
\begin{figure}[!h]
\begin{center}
\includegraphics[width=0.45\textwidth]{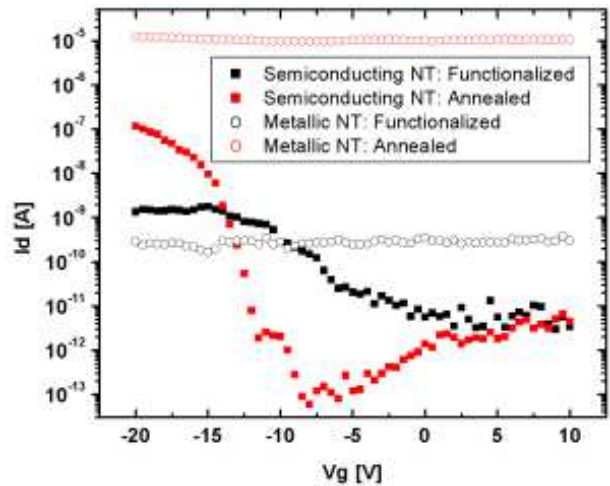}
\caption{\it The transfer characteristic Id-Vg using functionalized carbon nanotubes on a 100~nm SiO2 substrate before (black) and after (red) annealing (Vds = -0.5 V). The annealing improves the electrical contact between the nanotubes and the contact metal, and simultaneously the scattering introduced by the covalent functionalization is reduced.}
\label{F03}
\end{center}
\end{figure}
%\end{wrapfigure}

A key requirement of any device fabrication process based on chemical functionalization is that it does not adversely impact electrical performance. A strong indication for the loss of the functionalization and recovery of the lattice properties of the nanotubes after annealing can be seen in electrical transport measurements. The functionalized nanotubes were spread on a substrate with highly doped Si as gate, 100~nm SiO2 as gate dielectric, and 2~nm of oxidized Al as adhesion promoter for the functionalized nanotubes. Finally, the nanotubes were contacted with 5~\AA~Ti and 400~\AA~Au. Fig.~\ref{F03} shows a typical transfer characteristic of functionalized metallic and semiconducting CNTFETs. The devices show relatively low currents and poor performance. However, after annealing at 500$^{\circ}$C for 60~min the ON current of semiconducting devices improved by up to two orders of magnitude and metallic devices improved by as much as five orders of magnitude. Out of 12 measured devices, 3 of 3 semiconducting devices showed an improvements of more than one order of magnitude, and 6 of 9 metallic devices showed an improvement of four orders of magnitude or more. There could be two reasons for this behavior. First, the metal-nanotube contacts improve after the annealing due to the higher diffusion mobility of the metal atoms at elevated temperatures. Second, the covalent functionalization introduces scattering~\cite{C34} for charges propagating in the nanotube. This scattering decreases the ON-current and decreases the inverse sub-threshold slope. After annealing the functionalization disappears and the electronic structure and the good electric properties of CNTs are restored.

%\begin{wrapfigure}{r}{0.4\textwidth}
\begin{figure}[!h]
\begin{center}
\includegraphics[width=0.45\textwidth]{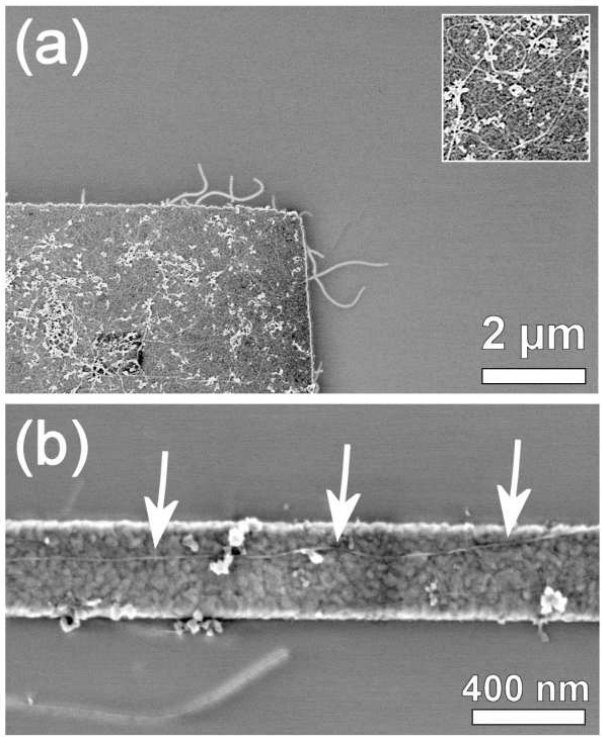}
\caption{\it SEM micrographs demonstrating (a) the selective deposition of functionalized nanotubes on Al2O3 (inset: magnified deposition region), and (b) that the nanotubes show a tendency to align on narrow features.}
\label{F04}
\end{center}
\end{figure}
%\end{wrapfigure}

The main goal of our functionalization is to selectively place nanotubes on a substrate. To test the selectivity we deposited the functionalized tubes dispersed in methanol on a SiO2 substrate patterned with Al2O3. The Al2O3 features were created by oxidizing Al films using oxygen plasma, yielding an oxide thickness of $\sim$4~nm. The nanotubes were deposited by drying a liquid film on such a sample on a hot plate at 100$^{\circ}$C. Then we sonicated the samples in pure methanol for 10~s followed by drying in N2. Fig.~\ref{F04}a shows a scanning electron microscope (SEM) image of patterned Al2O3 (left lower corner) on SiO2. The alumina has a dense nanotube layer. Some of the tubes branch out on the silicon oxide substrate (white fringes). The silicon oxide does not show any adsorbed nanotubes, indicating excellent site-specific bonding for our functionalized tubes. This was also corroborated by AFM. On narrow Al2O3 stripes aligned nanotubes can be observed (such as in Fig.~\ref{F04}b). Control experiments using unfunctionalized nanotubes showed no selective binding to Al2O3.  

%\begin{wrapfigure}{r}{0.4\textwidth}
\begin{figure}[!h]
\begin{center}
\includegraphics[width=0.45\textwidth]{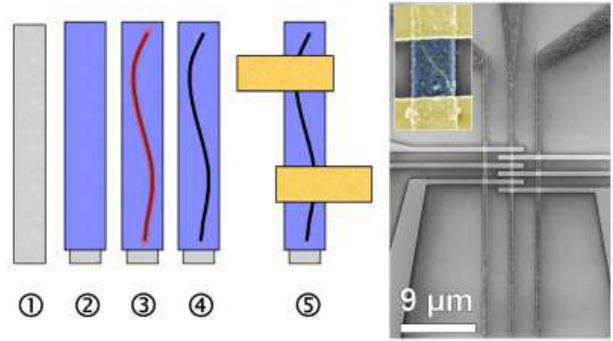}
\caption{\it Left: Schematic of the CNTFET assembly (process flow): (1) Definition of Al stripes, (2) Superficial oxidation renders an Al/Al2O3 structure, (3) Adsorption of functionalized nanotubes on Al2O3, (4) Annealing to remove the functionalization, (5) Contacting of nanotube with source and drain leads. Right: SEM micrograph of an actual device: The vertical leads are Al/Al2O3 structure with nanotubes adsorbed on top. The horizontal Pd leads contact individual carbon nanotubes on top of the gate structure. Inset: Junction with nanotube.}
\label{F05}
\end{center}
\end{figure}
%\end{wrapfigure}

We used the selective binding of the functionalized nanotubes to fabricate FETs with high yield. The approach is indicated schematically in Fig.~\ref{F05}. We first patterned a SiO2 substrate with 40~nm high and 300~nm wide Al stripes using electron beam lithography (1). Next, the samples were submitted to oxygen plasma (3~min at 600~mTorr) in order to oxidize the surface of the Al. Hence, inside of the structure the Al remains metallic and is protected from further oxidation by the surrounding oxide layer. This dielectric layer separates the Al gate metal and the semiconducting channel, i.e. the functionalized nanotubes (2). The functionalized nanotubes were then deposited as described above, resulting in nanotubes bonded to the Al2O3 gate dielectric, but not the surrounding SiO2 (3). The subsequent annealing in N2 removed the functionalization (4). Finally, the devices are completed in another lithography step in which 40~nm high Pd leads are deposited perpendicular to the Al/Al2O3 stripes (5). The FET channel length is about 400~nm. 

%\begin{wrapfigure}{r}{0.4\textwidth}
\begin{figure}[!h]
\begin{center}
\includegraphics[width=0.45\textwidth]{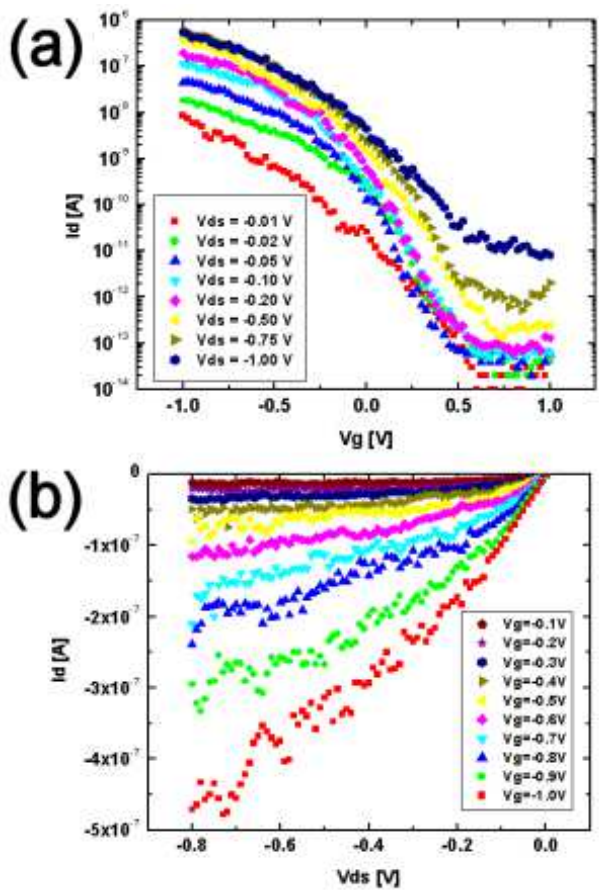}
\caption{\it (a) Transistors transfer characteristic Id-Vg of a CNTFET at different source-drain voltages, (b) Transistors output characteristic Id-Vds for different gate voltages.}
\label{F06}
\end{center}
\end{figure}
%\end{wrapfigure}

Fig.~\ref{F06}a shows the transfer characteristic of such an assembled CNTFET at different source-drain voltages. In this range of biases, the leakage current and the hysteresis are both small, which reflects the high quality of the gate oxide. With an ON current of more than 10$^{-7}$~A, an ON/OFF ratio of more than six orders of magnitude, a leakage current below 10$^{-12}$~A, a gate voltage switching range of $\pm$0.5 V, and an inverse sub-threshold slope of about 115 mV/dec at Vds = -0.2~V, such a device represents a high performance CNTFET with excellent properties. The good switching properties are attributed to the good electrical properties of the CNTs and the thin gate oxide layer~\cite{C35}. The increase of the OFF state current with more negative source-drain voltages is due to the thin gate oxide layer~\cite{C36}. Source-drain voltages more negative than -0.5~V give rise to minority-carrier injection at the drain leading to higher OFF currents. The output characteristic in Fig.~\ref{F06}b shows that the current turns on at Vds = 0~V and saturates early at a high level, indicating that the current is not limited by contact resistances or dominated by short channel effects. The device performance is comparable to that of pristine laser ablation nanotube devices. The device yield was also high. Out of 49 measured junctions, we found 28 working devices: 16 with semiconducting and 12 with metallic characteristic.
 
Using thin gate dielectrics leads to higher performances of field effect transistors~\cite{C37}. Additionally the use of Al2O3 increases the performance due to its higher dielectric constant compared to SiO2 typically used with CNTFETs. Our approach combines both advantages together with the selective large scale assembly. The device geometry allows addressing the devices individually. Further treatments such as moderate annealing~\cite{C38,C39}, or chemical doping~\cite{C40,C41} could improve the properties or change the characteristics of the CNTFETs. The availability of suspensions of only metallic or only semiconducting nanotubes would allow the fabrication of complex nanotube-based integrated structures. Exploiting the burning of nanotubes at undesired locations~\cite{C42} can lead to programmable logic and memory devices. Furthermore, our deposition method can easily be adapted to other basic metal oxides such as ZrO2, TiO2, or HfO2. Those with high dielectric constants are of particular technological interest.

In conclusion, we successfully functionalized carbon nanotubes using long-chained aryldiazonium salts end-capped with hydroxamic acid. This allows CNTs to be individually suspended in methanol or ethanol, while the hydroxamic acid group allows the selective deposition of the functionalized CNTs on Al/Al2O3 structures. An annealing step restores the structure and excellent electrical properties of pristine CNTs. The self-assembly into devices structures leads to high performance field effect transistors and provides the possibility of large scale integration.

\section*{Acknowledgement}

The authors thank Zhihong Chen for helpful discussions. Christian Klinke acknowledges financial support from the Alexander-von-Humboldt Foundation.

\clearpage

\end{document}